# Blockchain Synchronous Trust Consensus Model

Christopher Gorog MBA, PMP, CISSP, Terrance E. Boult IEEE Fellow
University of Colorado Colorado Springs

## 1 Abstract

This work introduces a novel approach for the governance of a blockchain containing social constructs and technical viability for widescale applications for the next generation of distributed ledgers. Functional requirements for this new blockchain distributed ledger (BDL) were garnered from an analysis of the needs for large scale applications. Applied research was employed as part of this endeavor to test the practicality and scalability of the solution outline. Novel features in this application draw together controls and enforcement for cybersecurity, digital content management, licensing, and configuration management. The Synchronous Trust Consensus Model™ applied research project named Project Philos™ was sponsored by the BlockChain Development Community (BCDC) with support from the University of Colorado[1], [2]. Research has followed both theorized conceptual and theory-to-practice models to prove the scientific soundness and the viability of incentive for community engagement. Results show that this new model proves the feasibility of an indefinitely expandable blockchain distributed ledger capability, while also providing a new participant incentive that is highly effective in engaging a community of practitioners.

## 2 Introduction

In this publication, we introduce a new type of consensus model which has included at its core a systems-engineering approach to blockchain distributed ledger. This approach was first theorized as a solution to design-limiting issues found during scale testing for blockchain applications. The proposed solution is now in the testing phase of an applied research project in partnership with the State of Colorado, University of Colorado Colorado Springs (UCCS), Colorado State University (CSU), and members of the Blockchain Development Community[1].

This model introduces several novel key features which include 1) indefinite transaction bandwidth and scaling capability, 2) trust-based incentive model eliminating waste from mining 3) integrated cybersecurity capabilities, 4) data management for end of life, and 5) a robust open governance model. With the systems engineering approach, several other features are addressed and additional capabilities added which were not previously envisioned with BDL.

It was imperative for this effort to elicit industry participation and acceptance which needed to include a widely engaging contributor incentivization such as those found with initially-successful cryptocurrency applications[3], [4]. The BlockChain Development Community, comprised of students from more than eight universities, and other corporate partners, was created to organize volunteers and form a public research organization charter as was found to be most effective by Sujan in 2018 aimed at embracing of new standards for market adoption[5]. Findings of Sujan, indicate that *"the relationship between the size of the venture and the market acceptance of an interface format is inverted U-shaped. This suggests that more members in the venture per se cannot be counted upon to lead to market acceptance of the format"*[5]. Leveraging these results, the BCDC charter initially engaged a limited number of contributing members but focused on building widescale awareness to potential user organizations. Also considering the findings of Son et al., 2019 is structured to support entrepreneurship in members and create a *"coopetitive environment"* as defined by Johansson et al., 2019 to elicit cooperation and competitive drive for its members[6], [7]. The BCDC charter was initially organized as a community-sourced project. In the future, the project is intended to be released as open source. Maintaining control with a manageable size group of contributors was deemed the best approach for the initial effort[5]. The success of the BCDC and partners have furthermore proven the exceptional value and enticement of the organizational structure and the project's inherent incentive structure.

## 3 Motivation

State of Colorado legislators have been enthused to find solutions to show accountability of State Government to constituents, and provide online privacy for Citizens[8]–[11]. The research for this publication was initiated to support such solutions for accountability and privacy using BDL. Early results of research to align with this direction saw a similar division of research topics as did Ademi, E., which were well classified as: *"1) blockchain reports, 2) blockchain improvement, and 3) blockchain application"*[12]. To move forward to the next phase and achieve an implementable solution, the decision was made to take a unique look at the strengths of BDL, and then add features to offset the limitations and/or weaknesses identified as the topics of many research publications[12]–[18].

Overcoming the limitations on blockchains would need to address a holistic solution which includes a systems-engineering approach and ground up design processes to attempt to resolve as many scalability-limiting issues as possible. Our review of these begins with the most scalability-limiting issues, the integration of best practices for security, as well as social and community needs for trust and visibility for participants. Yli-Huumo et al., 2016 report identified several limiting issues that remain true as this document is being amassed[13]. The inability to scale large programs using blockchain is twofold: the nature of the technical limitations, and due to the inability to control and govern an indefinitely expandable program using currently-implemented distributed ledgers.

# 4 Synchronous Trust Consensus Model™

We introduce the "Synchronous Trust Consensus Model™" (STCM) as a new direction for blockchain distributed ledger governance based on the results of applied research, combined with parallel features for best cybersecurity practices. The base concepts underlying blockchain distributed ledger are cryptologic operations put into practice using unique methods. We foresee that these technologies and industries will allow BDL to converge with other common cybersecurity practices[19], [20]. As the initial enthusiasm and glamour related to cryptocurrency wanes due to their functional limitations, exploring these connections are paramount for next-generation use cases[21]. Cryptography, being the underlying mathematical foundation for blockchain, and the primary method to implement cybersecurity, makes it only natural that blockchain research should be a primary consideration to discover and operationalize cybersecurity solutions. We have taken the approach,, as with the transitive property in mathematics, to consider Cybersecurity and Blockchain as irrevocably connected[22]. In this applied research we connected these practices as if they are a single technology and will eventually become integrated industries[19].

# 5 Background for This Work

The initial blockchain use case we envisioned was the distribution and logistics management of cryptographic keys for Internet of Things (IOT) devices[23]. The urgent need for scaling to meet demands for the number of computerized IOT devices worldwide – which is approaching hundreds of billions – required a new thought process for a blockchain that could ultimately scale to trillions of transactions. Early use cases provided an opportunity and support to study applications that required scaling to the scope needed for usage for a Colorado Statewide program[9], [24]. Findings of Ademi E. and Yli-Huumo et al., aligned with initial results from prototypes developed on a widely-adopted production blockchain, that project teams characterized as technology-limiting and completely inhibiting to the use of those target platforms[12], [13], [18], [25]. The results showed that scaling was not feasible using any current technology solutions evaluated. These initial findings were also instrumental in determining the functional needs, organizational requirements for scalability, and most importantly, garnering a foundation for additional research.

The initial test cases also identified that early BDL implementation lacked an ability to govern a social program in the constructs of the blockchain[26], [27]. Evaluation of statewide programs for the State of Colorado required constructs and capabilities related to social paradigms to support governance on a large scale[6], [24], [28]. Several examples of similar exploration can be found in many industries as governance across multiple industry participants is required. Examples can be found with the Pharmaceutical industry MediLedger Project[27], [29]. Since constructs required for governance are in a very early stage in platforms explored, it was determined for Project Philos™ that new constructs for governance needed to be built from the ground up. Our research concluded that governance capability would need to be integrated into the blockchain consensus constructs. Ultimately the new solution would contain a hybrid of successful traits found in various applications and additional functionality beneficial to the primary identified use cases[9], [10], [27].

Early research identified that the new direction would be most effective by associating a trust value related to actions performed, which is directly enforced within consensus itself. Additionally, trust relations of individual participants were identified as beneficial to enable the quality of transactions as a tool for governance of the overall blockchain. The inclusion of entity-related trust provides a major social constructthat can be used to resolve large globally unsolved problems in cybersecurity[30]. The categorization of trust and constructs of human trust assessment identified by Gorog, Boult in 2019 were incorporated into both the STCM and the novel Philos™ BDL design[23], [30]–[32].

## 5.1 Overcoming BDL Limitations

In a review of initial early blockchain concepts, it is apparent that social reform was the primary consideration and the application and supporting technology were less emphasized[3], [26], [31], [33]. The concept's success with Bitcoin showed that the consumer-base appetite existed for a digital-age trustless exchange capability[31]. Blockchain use beyond cryptocurrency has many practical and widescale uses. Over 71 distributed ledger program candidates have been identified within the state of Colorado[9], [34]. However, as initial research proved, existing blockchain technologies are not robust enough to support wide scale adoption[35], [36]. Limitations categorized in work from "Innovation and Software, Lappeenranta University of Technology, Lappeenranta, Finland," offered an accurate starting point for this work with the following categories of BDL limitation[13]: *"1) throughput, 2) latency, 3) size and bandwidth, 4) security, 5) wasted resources, 6) usability, versioning, hard forks, and multiple chains"*. The proposed Synchronous Trust Consensus Model™ (SCTM) and the implementation of the SCTM in the Philos™ Marketplace provide systems engineering solutions with workable features to solve all of these limitations. The application of components and parts of SCTM is summarized in the remainder of this article.

### 5.1.1 Indefinite Transaction Scalability

Our research identified an inability to scale as the principal limitation required to be overcome to meet the identified practical and social use cases[12], [18], [25], [36]–[38]. Many notable efforts are underway to solve scalability issues. Some have made successful strides, notably with the introduction of parallel chaining and side chain operations[17], [39]. By limiting the number of ledgers required to store duplicate transactions we sanction, over a given time a parallel side-chaining blockchain can handle more throughput then a serially restricted chain due to the nature of reduction of complexity. The STCM implements side chains by default to mirror successfully scalable models.

### 5.1.2 Growth of Distributed Ledgers Over Time

The initial intent and theory of operation for blockchain was to be able to trace all actions back to the first transaction held in the genesis block. Though this would accomplish the proposed social intent, it is apparent that this would not be able to support an indefinite number of participants making an indefinite number of transactions for an infinite period of time[16], [17]. The size of resources required to house the possible magnitude of transactions and the time required to verify each operation back to the genesis block reaches a point where the resources and time are not economically or practically feasible. The STCM implements a new feature *to* carry forward values and parameters with each *new* consensus block, removing the dependency on indefinite needs *to store* past transactions.

### 5.1.3 Storage Requirements

Data storage requirements for current distributed ledgers instances have already become problematic[16]. As sizes of overall ledgers grows, the growth in the number of transactions outpaces the storage needs supplied by people who choose to maintain independently-operated distributed ledgers[40]. This results in a dynamic imbalance between the needs for participation and the incentives offered by the technological implementation[41]. As times to verify transactions to the originating genesis block become unfeasible, third-party verification services become necessary to support existing implementations of blockchain[18]. The need for a verification service essentially negates the initial social motivation of independence from third-party trust described by Nakamoto, 2008[3].

The initial BDL models neglected to consider overall resource requirements, nor how they will scale with operations over time. The STCM's novel capabilities incorporates management and end-of-life planning for data to minimize adverse effects of scaling over time. It also adds a method for applying end-of-life criteria for blockchain data, as the reliance on legacy data is separated from operational criteria.

### 5.1.4 Longevity and Reduced Energy Footprint

Incentives for cryptocurrency which use transactional proof models have adversely increased the worldwide energy footprint. Leading research has identified this as an epidemic issue, as Sutherland, 2019 quoted[42] *"Blockchain applications must transition from proof-of-work toward energy-efficient consensus algorithms to be sustainable"*. Leading research has proposed that the trend or increased energy use by blockchains is in dire need to be reversed and that international treaties should be drafted to restrict blockchain mining[42], [43].

Another flaw with transactional proof-incentive models is they only support the initial phase of the overall transaction lifecycle[3], [4]. However, the needs for sustainability and longevity of the blockchain require transaction and data content availability for long periods of time. The entire conceptional basis of such incentive models conflict with best organizational practices summarized in, "On the Wisdom of Rewarding A While Hoping for B," as the expectation to maintain ledgers for an indefinite timeframe conflicts with incentives that only supports engagements during the initial transaction phase[44].

The STCM institutes a novel incentive structure which eliminates mining practices, and instead is based on participation over time. At the same time, the model is structured to align the reward with the needs to align longevity with best organizational practices. This new direction provides the groundwork for reversal of a global trend of energy waste. It is anticipated that this new direction can change the trend which is having a negative planetary wide impact[43], [45].

### 5.1.5 Data and Transaction Management

Most blockchain applications do not consider managing visibility of the data related to chain transactions at all. Our identified that the need for public blockchain content is limited to verification of the transactions, and also identified a need for private off-chain storage which could be verified by the chain transaction[4], [4], [46], [47]. The novel STCM approach addresses blockchain visibility and user privacy with dynamic user-controlled features.

A STCM user control feature allowing for the separation of transaction and off-chain content provides a social construct to enhance usability. This feature also provides a method to support regulatory compliance, jurisdictional control, and data management features for individual data owners to implement digital rights management[48].

## 5.2 Parallel Distribution of Ledgers

The STCM incorporates parallel capability using side chains as the base feature for the framework. The modularity introduced with the STCM to support scalability has been modeled after the internet's global expansive framework created by the router and domain name service (DNS)[49], [50]. The applied research Project Philos™ in partnership with BCDC has structured designs modeled after the internet and DNS service. The BCDC Project created two components: the "Peer Ledger" (PL) and the "Consortium

Server" (CS). Similar in features to the internet, the PL can be replicated indefinitely with each connection to the framework registered with a consortium service[49]. Once registered, a new PL enters the synchronous consensus and can continue to participate, while gaining incentive rewards incrementally over time. Each PL builds a reputation for performance and strengthens their standing as they build their trust ratings over time.

This overall system-of-systems creates what is more accurately characterized as a blockchain marketplace. Multiple use cases can coexist on this marketplace simultaneously without interfering with or affecting the performance of each other. Scaling both new use cases and the number of transactions required for an operational application are easily done by the addition of more PL systems. As the marketplace expands, the addition of consortium servers can also be scaled in a similar method by adding more systems.

### 5.2.1 Side Chaining by Design

In the STCM, each individual instance of a distributed PL creates its own blocks on an individualized side chain. These side chains created by each independent PLs come together in varying numbers with different requirements for consensus. This approach uses a similar method to the quorum slicing successful in the Stellar model[39]. The systematic functionality of the entire blockchain exhibits a semi-autonomous operation capability, modular in nature, while containing a global control structure to implement consensus. This division of the overall marketplace into side chains creates the ability to rapidly produce content block throughput. This component was implemented to address the limitations found on throughput of existing BDLs. The only restriction to content creation becomes the capabilities of the systems producing each side chain[13].

### 5.2.2 Asynchronous and Synchronous in Tandem

Human initiated operation exhibiting a random aspect to their origination. While many computer era operations are predictably timed and require extremely rapid responses. Initial applications of block chain only considered the asynchronous operations to support the human related transactions. Connecting cybersecurity best practices within the STCM model led to the addition of features which support rapid transactions and predictably timed operations to sustain the synchronous operation and periodicity of machines[30], [33], [51], [52]. To support connection of both human actions and machine periodicity, a novel feature is included as a base control in the STCM which alternates synchronous and asynchronous consensus levels. Asynchronous individual user actions originated by humans make content blocks which are inserted at the asynchronous sequence in which they are initiated. Synchronous requirements then restrict that consensus be made on time before the side chain creation of independent intervals content blocks can once again resume.

As successful in the Stellar model STCM is divided into three tiers much like the Leif, Mid and Top tier structure[39]. In the STCM these are Primary, Bridge and Global. Each tier is independently responsible for its own calculated values based on content created. Primary and Global consensus required operations are performed on a finite time interval.

In order to combine the primary tier and the and the top global tier which are both synchronous, an asynchronous middle bridging tier was incorporated into the model. This bridging tier serves a few needs including the ability to connect otherwise disconnected primary tier sidechain operations for value exchange. Operations of the bridging tier also include contributions in resolving double-spend attempts. Double-spend addressed by the Nakamoto and further researched by Rosenthal, embody attempts by participants to game the system by trying to participate in multiple transactions concurrently where they benefit at the expense of other participants[3], [53]. The STCM asynchronous bridge consensus uses larger quorum slices which verify the lower tier to provide higher certainty protection against double spend. Thought double spend is not fully resolved until transactions reach top tier (global) consensus, bridging confirms calculation for each lower tier enabling forensic inclusion of each side chain through verification by a larger percentage of ledgers[39].

### 5.2.3 Periodic and Cyclic Consensus

The primary grouping of PLs which produce a side chain is called a Primary Synchronization List (Sync List) or (PSL). This minimal grouping of PLs is a set of distributed blockchain PLs which stores redundant copies of the entire content of each of the members' side chain of blocks. The sync list enters into the primary consensus, creating a single consensus block containing the hash of the last block from each side chain on a reoccurring increment called the Primary Synchronization Interval. This feature addresses latency by reducing the requirements of data storage on duplicate ledgers to a manageable size. It also adds the first features of the STCM which provide for cybersecurity integration and trust building. The requirement for cyclic consensus incorporates the synchronous component into the STCM, departing from legacy BDL by the additional ability to relate enforcement of finite timing operations.

### 5.3 Intersection with Cybersecurity

With the ground-up applied research approach, cyber security best practice was incorporated into the STCM and Project Philos™ to enable a system-of-systems governance structure. The concept and features to support both layered defense-in-depth, and zero-trust were combined into the communications for each PL system[54]–[57]. Project Philos™ included a zero-trust wrapper and capability which restricts each system operationally to a uniform base structure. In cybersecurity terms this enables a configuration management structure built into the functionality of each distributed ledger[58], [59]. The configuration of each PL is verified and placed into immutable consensus blocks. In combination with the synchronous feature, this enables the enforcement of system health for each ledger as a marketplace-wide control[59].

### 5.3.1 Security and Privacy

With the integration of timing-enforced zero-trust, each PL system contains defense-in-depth with features similar to those found with multi-factor authentication[60]. The capabilities for zero-trust are enabled by blockchain-tracked provisioning of trust components in each PL[23], [54], [56], [61]. The net effect is that each system must be able to identify itself on initial communications with others in order to participate as a trusted PL node. This feature enables a largely missing capability in the digital age enabling a widespread ability to enforce compliance to a uniform set of standards.

### 5.3.2 Compliance Controls

The STCM novel security tools provide a push-based adherence capability for compliance enforcement. The nature of periodicity found in the STCM require strict operation on a periodic basis to permit PLs to be included into consensus. This control capability is used to enforce widespread compliance not only for the PL systems, but foreseeably also for many end use compliance applications.

Since the Project Philos™ primary objective was supporting governance visibility and meeting needs for State control, our research identified attempting to continuously identify rogue or ill-behaved nodes an extremely high risk. In order to mitigate this risk, a distributed ledger code verification feature was incorporated. This feature adds an additional layer of protection over Byzantine Fault Tolerance, Federated Byzantine Agreement, or other fault-tolerant consensus only[39], [62], [63].

A large focus of the Stellar model and a foreseen weakness is the ambiguity of codebase and lack of configuration control. This requires excessive effort to identify ill-behaved and well-behaved nodes[39]. The STCM version verification incentivizes community self-patrolling though the vested interest of each peer. Partner ledgers are penalized jointly and thus have incentive to verify partnering ledger configuration and compliance. The ability to control revisions over the entire BDL marketplace should also greatly reduce the capability for hard forks of the chain, addressing a major security risk inherent to most BDL applications. Our predictions are that several security and usability issues such as selfish mining, and majority attacks also may be mitigated or nearly eliminated with this control[14], [64].

To mitigate any risk associated with the STCM source control, the entire process is designed to be openly visible and neutrally-governed. A consortium governing body was formulated to maintain visibility to the revisioning and compliance structure[6]. This incentivization anticipates the evolution of a cooperative environment to foster expansion and adoption of the distributed ledger marketplace[28].

## 6 Operation of the STCM

The Synchronous Trust Consensus Model™ reduced to its most basic components consists of a Peer Ledger (PL) and Consortium Server (CS). The PL is a single autonomous system developed to operate with other PLs which operate on the same protocols and functionality as there own. Each distributed PL is a modular unit like the system of routers which comprise the Internet[49], [50]. A new PL can be entered into operation and registered with a local Consortium Server (CS) which then allows it to participate in unison with the system-of-systems created by the marketplace of all the registered PLs which maintain good standing. All the PL use the support of CSs' to continuously participate in the synchronous consensus.

### 6.1 Three Tier Consensus

Three tiers of consensus, Primary, Bridge, and Global, have different features and functional necessities in their support for achieving the STCM indefinite scalability. The content blocks, which correspond to Leif Tiers in other blockchain models, deliver the feature allowing rapidly-producible side chaining[39]. Throughput for production of content blocks, as well as the number of unrelated uses cases, is expandable by adding more lower-level tiers into the marketplace of distributed ledgers. The three-tier design is displayed in Figure "Three Levels of Consensus."

*Figure: Three Levels of Consensus*

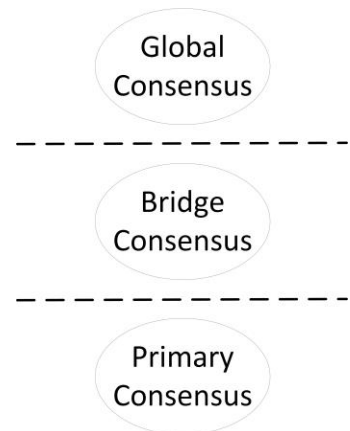

#### 6.1.1 Primary Consensus

The lowest tier of consensus is the Primary Consensus (P*), which supports uniformity of a subset of distributed ledgers. It is also the first level which contains a component to enable the primary participation incentive structure. Primary Consensus operations implement the lowest synchronous structured level which becomes the basis of trust calculations related to the performance of each peer ledger's adherence to synchronous requirements. The P* requires a one hundred percent successful consensus of each member of the Primary Synchronization List.

#### 6.1.2 Bridge Consensus

The middle tier consensus is the Bridge Consensus (B*) which supports the marketplace-wide impartial verification of subsets of P* operations. The B* is entered into by multiple groupings of synchronization lists and is done asynchronously when required to perform exchange*s* of value, renewal of transaction licenses, or establishment of global parameters. The success of B* depends on receiving a percentage of outstanding ledgers performing matching calculations which verify the results of calculations produced by each independent Synchronization List participating in the B*. The minimum percentage or ledgers required to create a B* is set by a governance organization's global configuration setting.

Though B* are done on an as-required basis, global required configuration settings enforce a maximum timespan between successful B*. When the maximum timespan is reached, the Synchronization List cannot continue creating side chains nor P* until they have performed a B* operation. The maximum timespan between B* is set by a governance organization's global configuration setting. This requirement also ensures the systematic participation in global consensus by the PSL, which may be offline or otherwise restricted from participating in a global consensus. Even with the occurrence of a PSL missing global consensus, all B* performed between Global Consensus instances ensure that the prior PSL side chain content blocks are included in the Global Consensus.

### 6.1.3 Global Consensus

The top tier consensus named the Global Consensus (G*) provides a reoccurring requirement for a majority of the outstanding distributed ledgers to reach consensus. This serves to eliminate any possibility of double spend across the entire marketplace, provides the global compliance control feature, and enables the global governance capability.

Once a day at the predetermined time set by the governance organization in a global configuration parameter, all active ledgers produce a final set of B* operations, which are then committed globally for consensus. All active root-level Consortium Servers, and a majority percentage of active PLs, are required to participate in the election process and create each G*.

In the occurrence that a PSL is offline or otherwise unable to participate in G*, the last successful B* ensures that content blocks created prior to that B* are included in the G*. An exception to this is the requirement that PSL participants are only permitted to forgo G* a limited number of times. This permissible number of missed G* is set by a governance organization's global configuration value setting.

### 6.1.4 Global Consortium Services

Along with the Peer Ledger, the second component which makes up the expansible distributed ledger marketplace is the Consortium Server. A Consortium Server is a single autonomous system also, like the Peer Ledgers, developed to operate together with other CSs identical to itself in order to provide a global service. The Consortium service provided by the collection of CSs supports the coordination and governance of all the PLs participating in the entire distributed ledger marketplace. Again, in comparison to the Internet, the CS can be associated with the type of services found in the Domain Naming System and Domain Name Server (DNS) support[49], [50].

It would be improbable for any design to encompass all considerations, thus allowance for future feature discovery was anticipated in the strategy. To this end, global controls for parameters are provided by the Consortium service. Each parameter is available on a pull basis, with the responsibility of each participating system to retrieve them. Consortium administrative controls allows for the augmentation of current parameters, or incorporation of unidentified future parameters.

Consortium administrative console workflow provides a structure for the governing body to propose the addition or augmentation of global parameters or permissions. The structure incorporates a public method of displaying proposed parameters, prior to promoting them to global use. Once the parameter is proposed and publicly vetted the final workflow stage places the parameter into immutable G*. This final stage places parameters into side chains reflected on all Consortium Servers, where they are available to each PL and globally enforced through the compliance control structure.

### 6.1.5 Management for Expandability

Our research and the applied Project Philos™ identified the beneficial application of smart contracts as an indispensable feature for the application of the disturbed ledger marketplace[4], [26]. However, a modular approach with minimal uniform block size was desirable to support rapid large-scale operations and facilitate the STCM incentive model. A novel feature was incorporated providing the separation of smart contracts from the blockchain content. Smart contract operations were incorporated into the STCM but use an innovative structure called Transactions Contracts (TC).

Instead of placing the entirety of the TC on the chain, its prototype definition is maintained as a global parameter. The publicly displayed TC prototype definition incorporates the contract logic only devoid of any individualized transaction content. The TC is registered as a part of the consortium configuration process, which is displayed publicly for vetting. Each TC prototype must be entered into immutable G* similarly to global configuration parameters before they are validated for use. Once registered, they are then available to the owner to use, or could be licensed to a third party. A PL uses a single TC repeatedly to produce the resultant output, which becomes the data within each content block inserted into its side chain. This process provides for expandability while supporting minimal content block size.

Features of the TC enables self-defined validation capability of content, as well as a bility for identification of private and public components. Execution of the TC 1) performs any self-validation of content, 2) separates content based on privacy classification, and 3) parses the content into either the side chain content block or offline data storage respectively. A properly defined TC will be publicly provable based on its prototype definition, and self-operating to achieve the restricted conditions of the contract provider. Signatures of the TC currently in use are placed into each content block created by the TC, providing verification of versions for each TC execution.

Needs for the Colorado Department of Regulatory Agencies (DORA) revealed that a direct alignment of license fulfillment as part of the immutable consensus would be beneficial. To meet this a novel approach to TC fulfillment was incorporated into Project Philos™. Similarly to the use of Gas within the Ethereum blockchain, licenses are expunged during the application of a Transaction Contract[4]. The project implemented a single license unit related to one instance of a content block created by the execution of a TC.

Each PL contains a license value, which represents the number of executions and subsequent blocks they are permitted to create from the execution of its registered TC. License units are pre-purchased from the TC owner and the units are expunged when used. The PL license value is decremented for each content block created. This structure provides a simple modular approach to purchasing, selling, and expunging transactions.

A biproduct of the TC decrementing license model provides for the pre-sale and modular management of licenses by TC owners. This new social construct provides for an entirely new licensing capability with wide scale applications[34], [65]. Whether the issuing agent is a DORA organization related to a government, a company creating legal forms, or an individual marketing digital rights for sale of unique content, any presale of digital content per transaction instance is enforceable. The vending of licenses by the owner to sub-issuers is also feasible by presale of license instances to third-party TC users.

## 6.2 Calculation of Trust and Credit

The STCM is not designed as a tokenized cryptocurrency; however, it contains a credit structure used within and as part of the consensus model. These credits within the Philos™ marketplace are called Phyli™. The Phyli™ structure is designed to be stable and not float in relation to market forces. To achieve stability, the Phyli™ is closely tied to the exercise of single transactions. The incentive structure provides a reward of a single Phyli™ for the storage of a single transaction. To share transactions with synchronized partners required the transferring of an equivalent number of Phyli™ to each partnering ledger.

### 6.2.1 Phyli™ Usage

As part of a consensus, each of the distributed nodes implements a net-zero calculation for exchange of Phyli™, based on the number of blocks each has made during the synchronization period. This final calculated value is established by each node and is recorded into each immutable consensus block. The primary consensus calculation is depicted in Figure "Net Zero Calculations." Each PL net-zero calculation at the top is based on one Phyli™ received for each block it receives to store from a PSL partner, minus one Phyli™ paid to each PSL partner for each block created and sent to that partner for storage.

Each consensus operation incorporates the net-zero calculation results for each PL augments the total outstanding Phyli™, and the new balance is written into the immutable consensus block. Depending on the consensus tier, requirements establish that all or an acceptable quorum of PL nodes must agree on each other's calculation before accepting that PL into consensus.

*Figure: Net Zero Calculations*

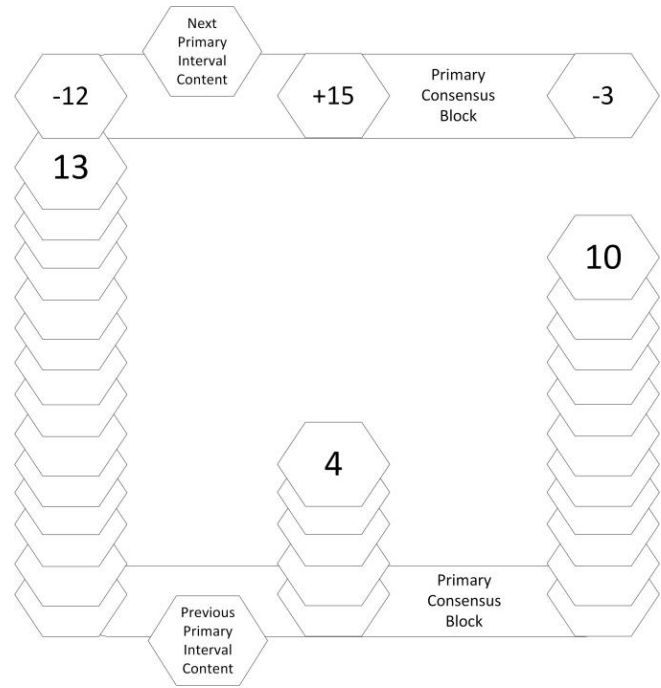

### 6.2.2 Trust Values

*One primary consensus interval is displayed with three participants in a Primary Synchronization List. Each of the three ledgers has produced sidechains with different numbers of blocks. The top Primary Consensus block shows the calculated net sum values for each of the three ledgers. These are the net value of Phyli™ either transferred to or received from primary synchronization list partners over the displayed primary synchronization interval.*

Another independent value for trust is maintained for each PL; the consensus operation also contains an algorithm to update this value each time a consensus operation is performed. The Trust value provides insight into the trustworthiness of the PL. This feature acts as a continuous evaluation of each PLs performance over time. The trust value embodies a similar relationship which is equated to consumer confidence, but it relates to each operating PL. This value is encoded directly into consensus blocks and recalculated each time consensus is done within the STCM. Each PL node correspondingly verifies the other's premeditated values by calculating the balance in the same manner as it does its own.

If any calculation of a Primary or Bridge is not excepted at a higher-level consensus, the content produced is not lost but must be restructured with accurate calculations after the erroneous tier successfully re-establishes synchronicity. The values of trust are established and recalculated based on the PLs ability to maintain synchronicity. Missing a consensus, for any reason, produces a reduction to the calculated trust value, which is re-adjusted upon the next successful consensus, and recorded in the immutable consensus block.

## 6.3 Building Influence

By default, the STCM design is "agnostic to" the inclusion of any single PL or CS but relies on the occurrence of a percentage of the active systems achieving the necessary sequence to meet consensus requirements. Should systems not meet sequencing, timing, or compliance requirements, they are instead removed from the roster of active systems, and the overall consensus continued after the reduction of participants. Individual participant ledgers' values grow with successful consensus participation, which determines their trust by performance over time. This produces a "Proof of Trust" nature to the overall consensus model, where the top performers are rewarded with more influence through increased trust scores.

In the STCM, there is no attempt to neutralize nodes that produce incorrect calculations as found in the Stellar model[39]. Since enforcement of configuration is performed globally, if all PLs perform correctly, they should operate on the same revision of the software specification/API. Should calculations not match between peer ledgers, objectives within the STCM shift to eliminating alliances with non-compliant ledgers. Quickly resolving non-compliance and restructuring to interfaces with compliant ledgers is a behavior that benefits participants. Thus coopetition is structured through Proof of Trust, where uniformity compliance encourages competitors to work toward a common objective. Trust visibility and association with the STCM incentive model are designed to influence the behavior of participants over time by incentivizing expected activities. As shown in multiple research studies, e.g., [28], [44], designing for the proper incentive is critical for long term effectiveness of a system

Maintaining synchronicity and active status in STCM is dependent on connectivity, reliability of PSL partners, and utilizing current global parameters. Maintaining revision integrity, and all configurations are the basis of the reward systems. Active participation with best practices for security of platforms and maintaining expected social behavior, increase trust and influence and avoids any individual penalization. The STCM marketplace, at its core, incentivizes uniformity. In theory, even in the occurrence of a system-wide flaw, which foreseeably could be caused by a defect within a configuration enforced revision, it is more likely that penalization would be uniform. Supersession of allowable versions and the processes for pre-vetting are handled by the neutral consortium governance body.

Primary use cases for the State of Colorado exhibited a high value for the accountability of participants and pre-authorizing and vetting transaction contributors. This requirement drove the need for the licensing feature which has shown comercial viability and favorable user community support. Responsible participation is emboldened by this feature as licenses of responsible entities retain greater quality, and thus value for selling and use by other parties. Licenses contain a transferable value, which is also increased through the process of building trust, empowering the participant with greater influence.

# 7 Results of Applied Research

The ongoing applied research Project Philos™ has displayed early success in several areas; the ability to engage public and private interest, the engagement of individuals with the proposed incentive model, and successfully meeting design objectives during early technology tests. The reduced size of each block enabled by use of the Transaction Contract and the Primary Synchronization List conception which reduces the number of ledgers required to store redundant content greatly impacted the STCM design.

Before initiating the concept for Project Philos™ team first evaluated Ethereum, Hyperledger Fabric, and Sawtooth Lake for possible candidates plaforms to implement Colorado State wide programs. Ethereum use was rulled out in early evaluation stages upon determination that the overall transactions for a relatively small State of Colorado program would produce 110,000 transactions per day from approximately 2,700 different end points. It was also assessed that larger scale programs candidates would require thousands of times more transactions. Sawtooth was elimated due to technology specific limitations, minimal documentation, and limited availability of support.

As Philos™ was designed to improved performance whild reducing the public chain footprint over equivalent distributed ledger applications. Considerations for this test examined the comparison to public content generated on chain between the evaluated platforms. Thought off chain data and/or private content would factor into growth, such content would be similar in operation and scale for all distribted ledger platforms thus was not considered for this evaluation.

The initial test platform was impelemented using Hyperledger Fabric in partnership with Colorado State University Pueblo. Hyperledger Fabric test application using a simple chain code was observed to grow the overall footprint in memory by four (4) Megabyte per each chain block. Though not tested at scale Ethereum was also assessed for projected growth and overall footprint, by equating the average smart contract size over the last two years of 22KB[66].

Testing was set up to compare an equivalent operation implemented on the BCDC Philos™ platform implementing STCM. The test simulated thirty (30) million transactions which mimicked an exponential increasing generation scale over three (3) years' time. Hyperledger Fabric was found to be able to keep up with the number of transactions over time but for all test casese the storage capacity used exceeded the capabilities of hardware implementing the test. It was assessed that this data set if it were implemented on hardware with enough storage resources would produce an overall growth of the chain to more than one hundred and fourteen (114) Terabytes in the first year of operation of the Statewide program.

The Philos™ implementation of STCM ran similar simulated scaling tests with the same thirty (30) million transactions. Test conditions were set to model the Primary Synchronization List containing three (3) peer ledgers where the test version of the chain block size was less than 400 bytes.

The resultant data set produced by Philos™ generated less than twelve (12) Gigabytes, accumulated only on the three distributed ledgers in the primary synchronization list. The resultant data generation for all other ledgers from the overhead of Bridge and Global consensus operations was less than two (2) Megabytes of content that was replicated on each distributed ledger.

In comparison to Ethereum size growth has been scaling over time as the Smart Contract gain more complexity[66]. Philos™ constant size of block gained by removing of the Transaction Contract from the public chain sets a linear growth as the chain scales. This together with reduced number of insances containing the entire chain, Philos™ show substantial scalabilitiy and perfance improvements providing alternatives for several identified Ethereum defincencies[38].

Thought the Philos™ prototype was in early stages during this test and the standard block size has increased in later versions it is projected to remain around 500 bytes in production released versions, incorporating no more than 10-20% change from current results. The test showed that performance and scalability under the STCM provided exceptional improvements over compared legacy blockchain applications. With the metrics garnered from these early tests the STCM exhibits an overall capability increase which would be manageable and able to be contanined within random access memory on an enterprise systems. The same scaling on Hyperledger Fabric would exceed the size of most disc drives for enterprise systems.

Growth as new PSLs are added to the STCM has the net effect of adding more parallel sidechains of independent groups of ledgers. Each of the PSL would perform in relatively the same manor with content creation only limited to the performance of the systems composing each PSL.

Overhead of the operations of Bridge will grow in a linear fashion with the growth of the numbers of PSLs in the overall marketplace. The growth is limited as operations producing linear growth include expansion from only each PSL group and not individual ledgers. Additionally, Bridge operations are done asynchronously such that the creation of content blocks by a PSL does not need to discontinue during Bridge operations. Once the bridge is complete the Primary Consensus operations would continue by re-defining the forensic hash linking of any content blocks created which are queued during the Bridge operation.

Global consensus operations are limited in growth as they only scale due to the growth in the magnitude of Bridges. SCTM limits this impact by continually calculating the Bridge content as they occur throughout the span of time between Global Consensuses. As with the Bridge operations, the Global consensus occurs as a background process supported by all of the Philos™ systems working in concert. Any content blocks created during this background process are also queued and forensically link by reassigning forensic hash chains after the Global Consensus is complete.

## 8 Challenges With STCM

Any new direction in technology while solving existing issues inherently may also lead to new challenges. Several foreseeable tradeoffs that have been identified and/or address within the STCM presented initial challenges. Though there is no way to ensure that every aspect will be proven effective or problematic without wider adoption, project teams are encouraged by the success received thus far. The following are some issues and tradeoffs that have been identified as challenges to the overall model, which may be mitigated as work continues.

### 8.1 Hybrid Limited- Decentralization

It is notable during early prototyping that in order to align the governance capability that several areas of the STCM may exhibit moderately centralized structures or controls. Exceptional considerations were made in order to implement workarounds by either the operation of redundant structures or by mitigation of the impact through openly visible governance of the respective features. However, these design tradeoffs may be assessed by some people to portray limited decentralization and may offer concerns for proponents of absolute decentralization. As with any technology, these features, which are essential to the needs of the primary use case*s,* were handled by offering consideration to mitigating risks in order to manage expectations of existing distributed ledger community users.

### 8.2 Searching and Visibility

As users of the technological era have become accustomed to the largely unrestricted access of data, the ability to control data content on this novel platform departs somewhat from the current expectation of "normal" data visibility to which many have become accustomed. The ability to globally search content and even the visibility of metadata in this new model is largely in the hands of individual data owners. Foreseeably this may take some time for user community and application use cases to fully embrace. During the evolution of this new model, it is anticipated that minimal visibility may result in many instances of orphaned content or a reduced awareness of data available for legitimate use. It is foreseeable that future use cases will adapt to this new paradigm, but also that undesirable results will persist from lack of visibility or search capability.

## 9 Identified Strengths of STCM

As Project Philos™ has unfolded, several unanticipated strengths have emerged. Many or these are unexpected or emergent behavior resulting from the effects of linking of cybersecurity with the distributed ledger. This list is only a portion of the major topic areas where use case discussions have identified entirely novel uses or application capabilities.

- Elimination of proof of work energy waste
- Patch management and version enforcement
- Privacy of owner-creator content
- Secondary data sale market
- Self-Sovereign identity[67]
- Private sector-issued licensing
- Zero-trust expanded application
- Provide risk and trust relationships
- Provide trust and quality relationships
- Insurer and banking models
- Disconnected governing and jurisdictional control
- Enforcement and audit by legitimate authorities
- Digital rights management
- Pre-paid models for security issuance
- Quantifiable metrics replace qualitative processes

## 10 Conclusion

We have proposed a novel consensus model and an indefinitely expandable approach to blockchain distributed ledger. We started with the pretense of taking a modular system-engineering approach and incorporating re-designs to correct known scalability limitations of legacy blockchain applications. Considerations were incorporated for best practices and theories of successful blockchain applications, while including features supporting state-of-the-art cybersecurity practices. Novel incentive models for robust community participation were conceived based on an analysis of organizational needs and desires for reduction of worldwide energy waste. This new model for blockchain is agnostic to the inclusion of any single ledger but incentivizes uniformity and continued participation over time. The novel use of Proof of Trust in a consortium-governed structure opens countless relational use cases. Initial results of prototype platforms show robust community engagement, and capability for widely scalable transaction growth. Metrics garnered from early prototype tests exhibits an overall impact of greatly reducing the system resource needs to operate an independent distributed ledger.